# LARGE GEOMAGNETIC STORMS ASSOCIATED WITH LIMB HALO CORONAL MASS EJECTIONS [*]

NAT GOPALSWAMY

*NASA Goddard Space Flight Center*
*Greenbelt, MD 20771, USA*

SEIJI YASHIRO[†], HONG XIE, SACHIKO AKIYAMA, and PERTTI MÄKELÄ

*The Catholic University of America, Washington DC 20064, USA*

Solar cycle 23 witnessed the observation of hundreds of halo coronal mass ejections (CMEs), thanks to the high dynamic range and extended field of view of the Large Angle and Spectrometric Coronagraph (LASCO) on board the Solar and Heliospheric Observatory (SOHO) mission. More than two thirds of halo CMEs originating on the front side of the Sun have been found to be geoeffective (Dst ≤ -50 nT). The delay time between the onset of halo CMEs and the peak of ensuing geomagnetic storms has been found to depend on the solar source location (Gopalswamy et al., 2007). In particular, limb halo CMEs (source longitude > 45º) have a 20% shorter delay time on the average. It was suggested that the geomagnetic storms due to limb halos must be due to the sheath portion of the interplanetary CMEs (ICMEs) so that the shorter delay time can be accounted for. We confirm this suggestion by examining the sheath and ejecta portions of ICMEs from Wind and ACE data that correspond to the limb halos. Detailed examination showed that three pairs of limb halos were interacting events. Geomagnetic storms following five limb halos were actually produced by other disk halos. The storms followed by four isolated limb halos and the ones associated with interacting limb halos, were all due to the sheath portions of ICMEs.

## 1. Introduction

Halo coronal mass ejections (CMEs) occurring on the frontside of the Sun are a potential source of geomagnetic storms because they can directly impact Earth's magnetosphere with high kinetic energy [1,2]. The geoeffectiveness of halo

---

[*] This work is supported by NASA LWS TR&T program.
[†] Also at Interferometrics, VA, USA.

CMEs depends on the existence of southward component of the magnetic field in the sheath and/or ejecta portions. Here we define geoeffectiveness as the ability of a CME to produce a geomagnetic storm with an intensity level measured by the Dst index at or below -50 nT, e.g., [3]. In a recent investigation of the geoeffectiveness of halo CMEs (Gopalswamy et al., 2007 [2], herein after Paper 1), it was shown that the geoeffectiveness declines as the source region of halo CMEs has a greater central meridian distance (CMD). It was also found that halo CMEs associated with intense geomagnetic storms (Dst $\leq -100$ nT) are generally located within a longitude range of $\pm 45°$ (average longitude ~W10) whereas non-geoeffective halos (Dst > -50 nT) had a broad longitude distribution ($\pm 90°$). Furthermore, ~75% of disk (CMD $\leq 45°$) halos were geoeffective while only 60% of the limb ($45° <$ CMD $\leq 90°$) halos were geoeffective. The computed the delay time between the CME onset at the Sun and the peak of the geomagnetic storm was surprisingly different on the average for limb halos (56 hr) and disk halos (70 hr). Paper 1 attributed this difference to the possibility that the sheath of the interplanetary (IP) CMEs (ICMEs) developing from limb halos must have produced the geomagnetic storms (sheaths are typically ahead of ICMEs by ~ half a day [4-6]). It is also known statistically (from ICME observations) that the sheath storms are generally ahead and the cloud storms are behind the arrival of ICMEs [7]. However, detailed investigation of the IP counterparts of individual halo CMEs and the associated geomagnetic storms was not made in Paper 1. The purpose of this paper is to provide a direct confirmation that the geomagnetic storms associated with limb halos are due to sheaths in the corresponding ICMEs. To this end, we examine the IP counterparts of the limb halos reported in Paper 1 to see if the sheaths of the ICMEs from limb halos have large southward magnetic field component to make them geoeffective.

## 2. Data Selection

Paper 1 listed 37 limb halos ($45° <$ CMD $\leq 90°$) that were followed by Dst values at or below -50 nT. The listed CMEs may overlap with other sources of geomagnetic storms, such as corotating interaction regions (CIRs) formed by high speed streams from coronal holes. It is well known that CIR storms generally have a Dst index $\geq -100$ nT [8]. To eliminate the possibility that some of the weaker storms may be caused by CIRs, we consider only strongly geoeffective limb halos (Dst $\leq -100$ nT). There were 17 such limb halos as listed in Table 1. The simple criterion for geoeffectiveness used in Paper 1 was that the halo CME must be followed by a geomagnetic storm during a 4-day interval starting one day after the CME onset. This criterion was based on the observation that it takes anywhere between 1 and 4 days for a CME to travel to Earth after the liftoff. One cannot avoid the situation that the time windows of CMEs overlap, especially during solar maximum when CMEs occur in quick succession from the same active region or from different active regions. This will result in some geomagnetic storms getting assigned to more than one CME: there may be a disk halo occurring around the time of a limb halo by chance, in which case one has to carefully decide which CME is responsible for the ensuing storm. We carefully examined all possible CMEs occurring around the time of the limb halos to determine whether it is truly geoeffective or not.

Table 1. List of limb halos followed by intense geomagnetic storms (1996 -2005)

| No | CME Date & Time | V km/s | Source Location | Dst Peak Time | DT hour | Dst (nT) | Notes |
|---|---|---|---|---|---|---|---|
| 1 | 00/04/04 16:32 | 1118 | N16W66 | 04/07 00 | 55.5 | -288 | Sh |
| 2 | 00/10/24 08:26 | 800 | S23E70 | 10/29 03 | - | -127 | CC |
| 3 | 00/10/25 08:26 | 770 | N09W63 | 10/29 03 | 90.5 | -127 | Sh |
| 4 | 00/11/25 01:31 | 2519 | N07E50 | 11/29 13 | - | -119 | CC |
| 5 | 01/10/01 05:30 | 1405 | S24W81 | 10/03 14 | - | -166 | CC |
| 6 | 01/11/22 20:30 | 1443 | S25W67 | 11/24 16 | 43.5 | -221 | INT |
| 7 | 02/03/22 11:06 | 1750 | S10W90 | 03/24 09 | - | -100 | CC |
| 8 | 03/06/15 23:54 | 2053 | S07E80 | 06/18 09 | 57.0 | -141 | Sh |
| 9 | 04/11/09 17:26 | 2000 | N08W51 | 11/10 19 | - | -131 | Rec |
| 10 | 04/11/10 02:26 | 3387 | N09W49 | 11/11 05 | - | -113 | Rec |
| 11 | 05/01/19 08:29 | 2020 | N15W51 | 01/22 06 | - | -105 | INT |
| 12 | 05/01/20 06:54 | 3242 | N14W61 | 01/22 06 | 47.0 | -105 | Sh |
| 13 | 05/05/11 20:13 | 550 | S11W51 | 05/15 08 | - | -263 | CC |
| 14 | 05/08/22 01:31 | 1194 | S11W54 | 08/24 11 | - | -216 | INT |
| 15 | 05/08/22 17:30 | 2378 | S13W65 | 08/24 11 | 41.5 | -216 | Sh |
| 16 | 05/08/23 14:54 | 1929 | S14W90 | 08/24 16 | - | -160 | Rec |
| 17 | 05/09/09 19:48 | 2257 | S12E67 | 09/11 10 | 38.0 | -147 | Sh |

Column 2 of Table 1 gives the starting date and time (yy/mm/dd hh:mm format) of the limb halos with their sky-plane speed (V in km/s) and heliographic location of the solar source taken from Paper 1. The time of minimum Dst of the associated storms is listed in column 5 in the mm/dd hh format (the year is the same as in column 2). The delay time (DT) from the CME onset (column 2) to the time of Dst minimum (column 5) is listed in column 6. The minimum value of the Dst index is given in column 7. Finally, some comments on the events are given in the last column (Sh – isolated sheath event; CC – chance coincidence; INT – interacting event; Rec – fluctuation in the recovery phase of a preceding storm).

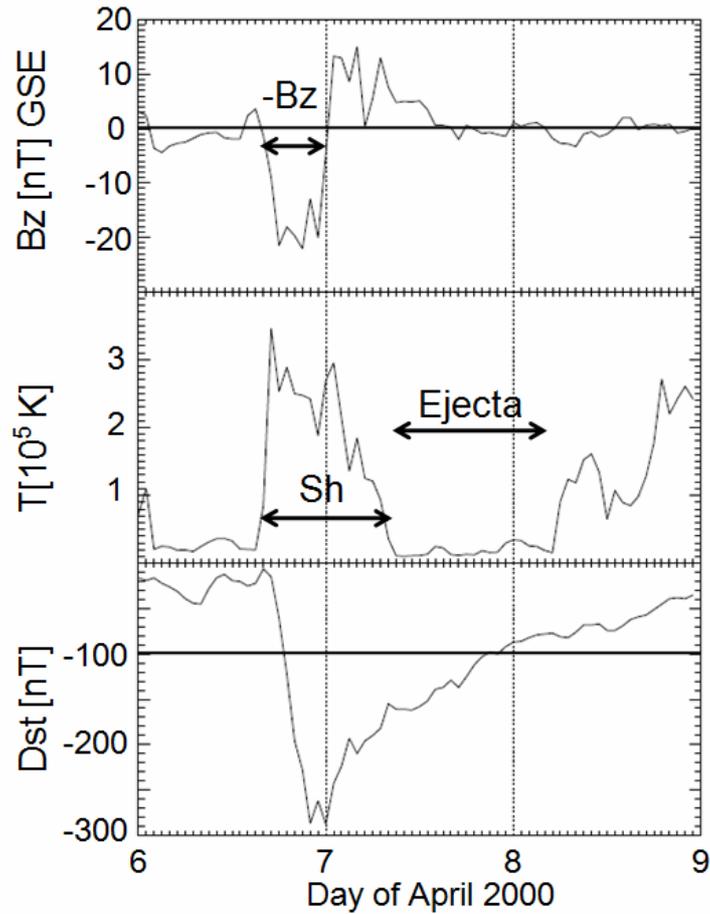

Figure 1. (top to bottom) Z-component of the IP magnetic field (Bz), solar wind proton temperature (T) and the Dst index around the time of the April 7, 2000 storm (due to halo #1). Intervals of Bz<0 (- Bz), sheath (Sh) and the ejecta (Ejecta) are marked. Note that Bz<0 occurs only in the front part of the sheath.

## 3. Analysis

Figure 1 shows the out of the ecliptic component (Bz) of the IP magnetic field (IMF), the solar wind plasma temperature (T) and the Dst index. From the temperature signature we can identify the sheath (marked Sh) and the ejecta (also marked). The ejecta is of short duration because the CME is not directed along the Sun-Earth line. Note that the intense geomagnetic storm is entirely due to the Bz<0 in the sheath region. The short duration ejecta has no Bz<0, so it is not geoeffective. Examining plots like the ones in Fig. 1, we found that CMEs #3, #8, #12, #15 and #17 all produced geomagnetic storms because of their sheath portions.

The halo CMEs #2 and #3 are both candidate sources of the same storm. Looking at the solar source, we see that the eastern source is at a larger distance from the disk center. Since CMEs are deflected to the east [9], we conclude that halo #3 is the likely candidate and regarded the association between halo #2 and the storm is by chance coincidence (CC). Halo #3 also resulted in an ejecta following the sheath. Bz<0 occurred in the sheath and partly in the ejecta, but the minimum Bz occurred in the sheath. We therefore, conclude that the storm is due to the sheath.

Halo #4 is very fast (2519 km/s), so the shock is expected to arrive in about a day. The shock actually arrives on November 26 at 11:40 UT followed by a narrow ejecta on November 27 at 12:30 UT. It is also associated with a moderate storm (~ -80 nT) due to its sheath but this is not the storm listed in Table 1. The storm listed in Table 1 is due to another CME on November 26 at 17:06 UT, which is a disk halo (N18W38). Therefore, we regard halo #4 to be a chance coincidence. Similarly halo #5 is a chance coincidence since the -166 nT storm is caused by the disk halo (N13E03) on 2001 September 29 at 11:54 UT. Halo #5 is also too close to the limb, which is unlikely to produce ejecta at Earth.

Halo #6 is followed within 3 h by another disk halo (S17W36) on 2001 November 22 at 23:30 UT. The two CMEs seem to have interacted near the Sun, so we cannot rule out the possibility that the sheath of halo #6 is swept up by the following disk halo. Therefore, we regard this as an interaction event (INT). The geomagnetic storm was caused by Bz < 0 in the sheath of the merged ICME at Earth.

The storm listed in the time window of halo #7 has a better candidate: the disk halo (S17W20) of 2002 March 20 at 17:54 UT, which had a shock and ejecta. The limb halo #7 did produce an IP shock that seems to pass through the ejecta from the disk halo. Therefore, we conclude that although halo #7 has an associated IP shock, it is not associated with the storm listed in its time window. The storm itself is caused by the ejecta part of the ICME associated with the disk halo.

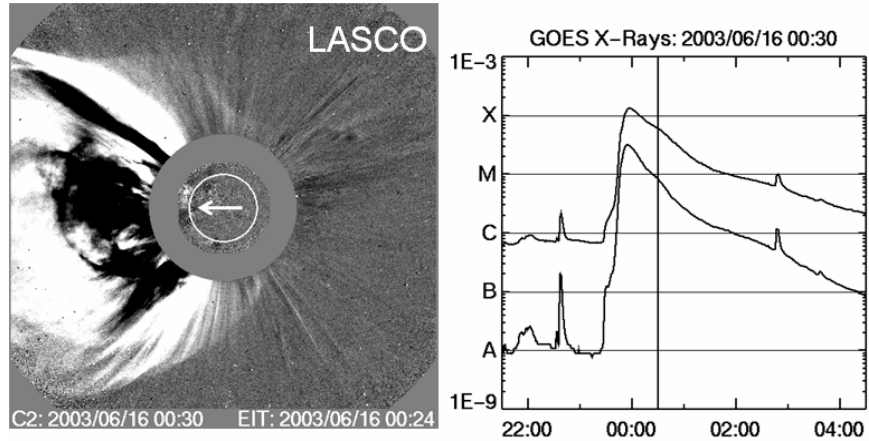

Figure 2. (left) White-light CME (halo #8) from SOHO/LASCO with superposed EUV difference image showing the solar source (pointed by the arrow). (right) GOES light curve showing the X-class flare associated with the CME.

Halo #8 is rather isolated and its association with the 2003 June 18 storm (Dst ~ - 141 nT) is unambiguous. In Fig. 2, we show halo #8 (2003 June 15 CME) in the LASCO frame obtained early on June 16. The solar source is clearly near the east limb (S07E80) as evidenced by the large-scale EUV disturbance and the associated X-class flare (see soft X-ray light curve from the GOES satellite in Fig. 2). It is clear from the LASCO frame that the western flank of the disturbance has crossed the Sun-Earth line early in the event. The CME was also associated with an intense type II radio burst in the decameter-hectometric (DH) wavelengths. The DH type II bursts are indicative of CME-driven shocks in the near-Sun IP medium. There were several small CMEs (widths ranging from $13^\circ$ to $40^\circ$) after the limb halo, but none of them is capable of producing a shock at 1 AU. The next significant event was a halo at the end of June 17, which was just 5 hours before the shock arrival at Earth and hence could not be the source. Halo #8 is also unique in that it is the easternmost CME to produce a major geomagnetic storm during solar cycle 23.

The solar wind plasma and magnetic signatures of halo #8 are shown in Fig. 3. The shock arrived at 04:44 UT on June 18, indicating a transit time of ~53 h. This is rather long for a 2053 km/s CME, but the Earthward speed is expected to be smaller because only the western flank of the shock seems to have arrived at Earth. The sheath that follows the shock is rather extended (more than one day).

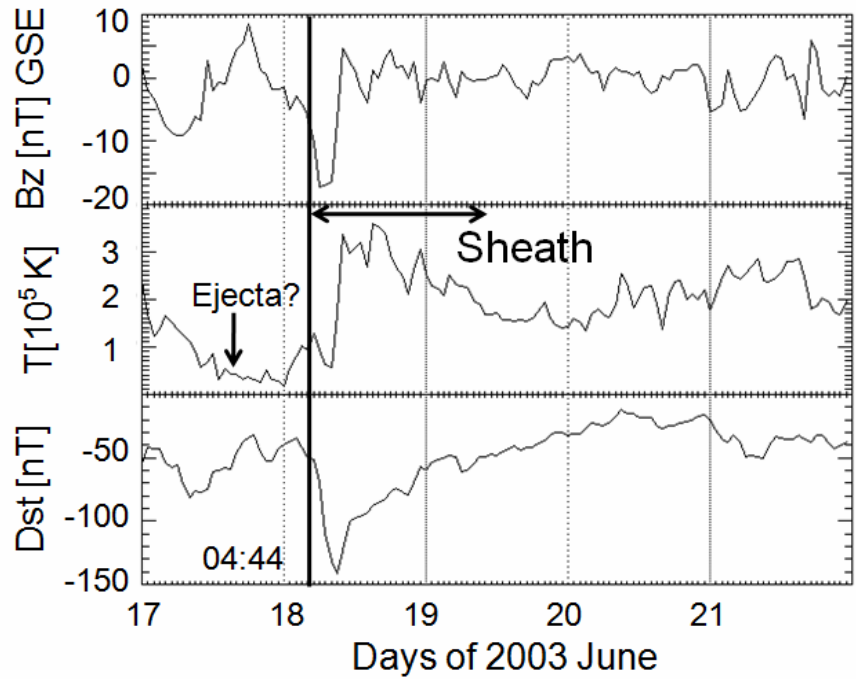

Figure 3. Solar wind magnetic and plasma signatures of the IP disturbance associated with the 2003 June 18 geomagnetic storm (followed by halo #8). The shock maybe running into a preceding ejecta (suggested by the depressed temperature) but there is no ejecta signature following the shock sheath.

The Bz plot shows that the interval of Bz < 0 occurs right after the shock, in the front end of the sheath. The Dst minimum occurs just 4 hours after the shock arrival, again corresponding to the front end of the sheath. There is no indication of an ejecta after the shock, because the source is far from the disk center. Thus we conclude that this is clearly a sheath storm.

The storms listed in the time windows of halos #9 and #10 seem to be fluctuations in the recovery phase of the previous super storm (- 289 nT on 2004 November 10 at 10:00 UT caused by the disk halo that left the Sun on November 7 at 16:54 UT). Examination of the solar wind plasma and magnetic signatures shows that there is no shock or ejecta around the times of these two storms. There are only small negative excursions in Bz corresponding to the two Dst minima in question.

The storm on 2005 January 22 is in the time window of halos #11 and #12. Figure 4 shows the two CMEs at their first appearance in the LASCO field of view. Both appeared as non-halos in the northwest quadrant and expanded to

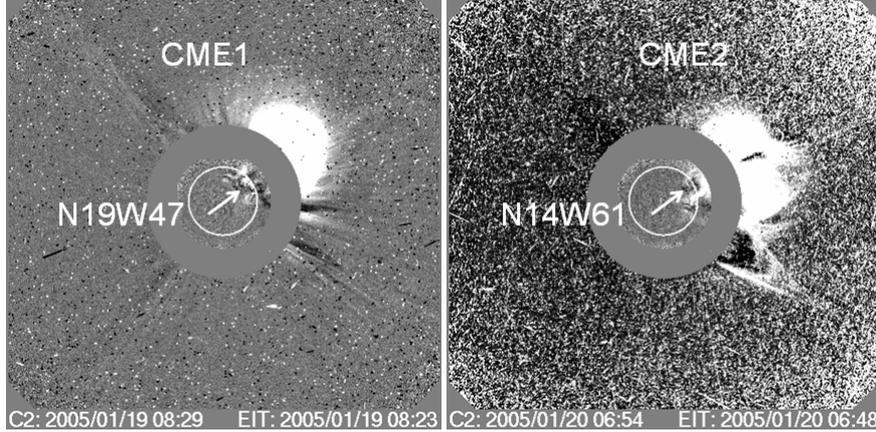

Figure 4. Two CMEs (halos #11 and #12 in Table 1) from the same active region (AR 0720) that become full halos in the LASCO/C3 field of view. The January 20 CME was visible only in a single LASCO frame because of degradation of the SOHO detectors due to impact by solar energetic particles from this CME [10]. The CME speed was estimated to be ~3242 km/s by combining the LASCO image with SOHO/EIT images that showed the eruption. Figure 5 shows the shock, the sheath, and the geomagnetic storm following the two halos. Note that Bz<0 occurs only for a short interval right after the shock at 16:48 UT on 2005 January 21. The sudden commencement in this case is extraordinarily intense with a positive excursion of ~30 nT. The two halos left the Sun within a time separation of ~23 h, so it is possible that the shocks from the halos merged to form the huge sudden commencement. The sheath shows a peculiar temperature structure (see Fig. 5), which may indicate that the sheath contains some portion of the ejecta of halo #11. There is a slight temperature depression after the sheath region, but there is no ejecta signature in Bz and By components of the IMF. This seems to be an interaction case although one cannot rule out the fact

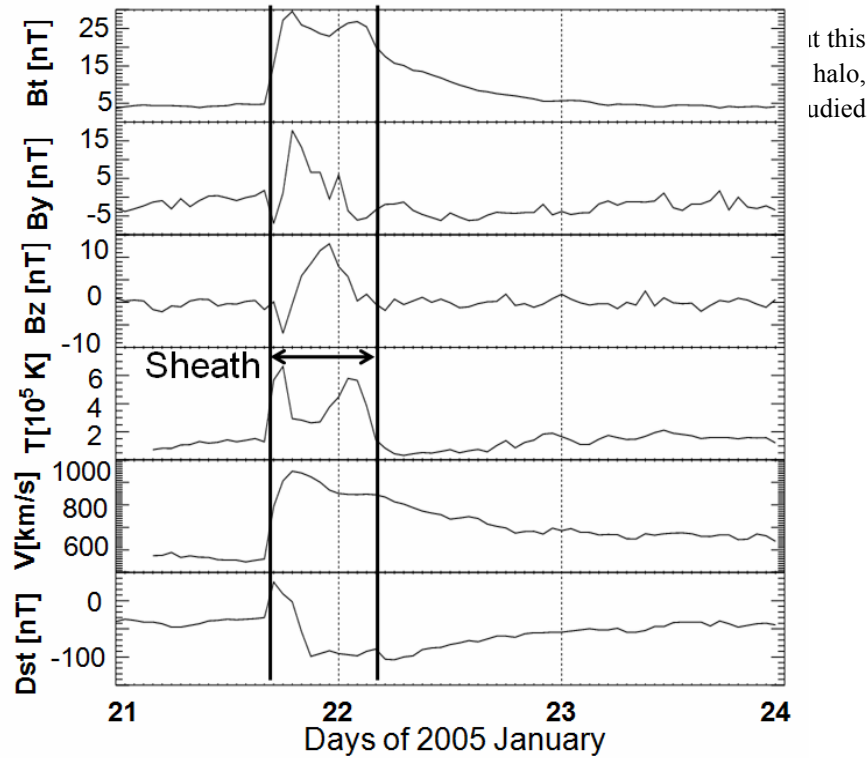

Figure 5. Magnetic and plasma signatures following halos #11 and #12 shown as magnetic field magnitude (Bt), the By component, the Bz component, the solar wind proton temperature (T) and flow speed (V), and the Dst index. The region of enhanced temperature marked as "sheath".

The storm on 2005 August 24 is in the time window of the limb halos #14 and #15, both of which occurred on August 22. There is only one IP shock observed at 1 AU (on August 24 at 5:34 UT). Right after the shock, Bz becomes negative and attains a large negative value (Bz ~ - 40 nT). The storm is due to this Bz<0 interval in the sheath region. It is possible that a second shock is present in the sheath region followed by a mini magnetic cloud (~2 h in duration) [13] but the proton temperature remains elevated as is normally the case in shock sheaths rather than during magnetic clouds. These two CMEs originated from an active region surrounded by a coronal hole, so the interaction seems to be complicated (see [13] for more details). Halo #15 occurred ~16 h after halo #14 and must have overtaken it somewhere between the Sun and Earth because both CMEs originated from the same active region. Halo #15 is twice as fast as halo #14, so the interaction is highly likely. However, the ejecta signature is not clear at 1 AU because the proton temperature remained above the pre-shock level. As in the case of the January 2005 events, the 2005 August 22 events were also interacting and resulted in a single shock at 1 AU. Again, we compute the delay time of the storm with respect to the first-appearance time of halo #15. The storm on 2005 August 24 at 16 UT is also a fluctuation in the recovery phase of the storm associated with halos #14 and #15. Even though the fluctuation appears in the time window of halo #16, we do not see any IP signatures of this CME. Note that halo #16 originated right at the west limb.

The last halo is one of the many halos from the super active region 0808 and one of the two superfast CMEs (speed > 2000 km/s). The IP shock associated with the CME was observed at the very beginning of September 11 (00:49 UT). The Bz turns negative right after the shock, well within the high proton temperature interval, so we are certain that the storm is due to the sheath region. The magnitude of Bz is not very high (~5 nT) but the speed is extremely high, so the storm is intense. Note that this is one of the smaller storms in Table 1.

Excluding the chance-coincidence cases (5) and the three recovery-phase fluctuations (3), we get 9 limb halos that were responsible for the 7 geoeffective intervals. In every single case, the storm was caused by the sheath of the IP counterparts of the halos, thus confirming the suggestion made in Paper 1. Since there were only 7 distinct storms that can be attributed to the limb halos, we have listed only 7 delay times (from CME onset to time of minimum Dst of the storm) in Table 1. For the three pairs of interacting CMEs, we counted only the faster, overtaking CME for computing the delay time. In one case, a disk halo was overtaking a limb halo, but the time difference was very small (~3 h). Four limb halos were isolated so there is no ambiguity in the delay time. The delay ranged from 38 h to 90.5 h, with an average value of 53.3 h, not too different from the average value (56 h) reported in Paper 1 for all geoeffective limb halos (including those associated with moderate storms).

## 4. Discussion

We studied the geoeffectiveness of 17 limb halo CMEs by examining their IP counterparts. In particular, we examined where the $Bz<0$ interval occurred: within the ICME interval and/or in the sheath ahead of the CME. In all the cases, in which we can make an unambiguous association between the limb halos and IP shocks, the geoeffectiveness is caused by the sheath ahead of the ICMEs. A suggestion to this effect was made in Paper 1 without examining the IP data. In this work, we have confirmed the suggestion by examining the solar wind plasma and magnetic signatures associated with the limb halo CMEs. When the ICMEs are shock driving, the sheath provides an additional source of $Bz<0$. If the ejecta part is a magnetic cloud, the $Bz<0$ interval can occur in the front or back of the cloud for bipolar clouds, throughout the cloud interval for south-pointing high-inclination clouds, and no interval of $Bz<0$ for north-pointing high-inclination clouds, e.g. [5]. For limb halos, the cloud part may or may not arrive at Earth; the sheath is likely to arrive at Earth and produce a geomagnetic storm if it has a $Bz<0$ interval. The lack of ejecta arrival at Earth reduces the probability of limb halos producing a storm, consistent with the central-to-limb variation of geoeffectiveness of halo CMEs reported in Paper 1.

The present study also confirms the delay time between the arrival of magnetic clouds and the time of minimum Dst during storms [5]. The average delay between sheath and cloud storms can be estimated from the fact that sheath storms are typically ~3h ahead of ICME arrival, while the cloud storms are ~11h behind the ICME arrival [7]. Thus the sheath storms are expected to be ~14 h ahead of cloud storms. For the set of events in Table 1, we arrived at an average delay time of ~53h, which is smaller than the value obtained for storms following disk halos by ~17 h.

One of the interesting outcomes of this study is that two of the four isolated limb halos are from close to the east limb (S07E80 for halo #8 and S12E67 for #17). This result is significant because it highlights the difficulty in forecasting geomagnetic storms based on CME observations. It is usually believed that CMEs occurring within ±30° from the disk center arrive at Earth and cause geomagnetic storms and that there is a slight western bias of the CME source regions on the Sun. Clearly CMEs originating close to the east limb also produce geomagnetic storms under extreme conditions (both the CMEs were superfast with speeds 2053km/s and 2257 km/s).

Another surprising result is that 5 of the 9 limb halos that resulted in large geomagnetic storms were interacting with other CMEs. In one case, the limb halo (#6) interacted with a disk halo. The remaining interactions were among limb halos (#11 with #12 and #14 with #15). All the five limb halos are known producers of type II radio bursts in the IP medium (http://cdaw.gsfc.nasa.gov/CME_list/radio/waves_type2.html). Type II radio bursts are indicative of CME-driven shocks because electrons accelerated at the shock front produce Langmuir waves, which in turn produce radio emission at the local plasma frequency or its harmonic. In other words, all the five halos drove shocks in the IP medium, but at 1 AU, each pair resulted in a single shock. This may mean either the shock of the preceding CME decayed or it merged with that of the second CME in the pair.

During the study period (1996-2005), there were 75 large geomagnetic storms (Dst < -100 nT) associated with CMEs [8]. It is interesting that 7 of them (or 9.3%) are due to limb halos.

## 5. Conclusions

By examining the IP counterparts of limb halo CMEs using solar wind plasma and magnetic signatures, we have confirmed that the geomagnetic storms following limb halos are caused by the southward component of the IP magnetic field contained in the ICME sheaths. Since the sheath is the first feature encountered by Earth's magnetosphere, the delay time between the onset of halo CMEs and the peak of ensuing geomagnetic storms is the smallest. The delay time is ~20% smaller for limb halos than for disk halos reported in Paper 1. We also confirm that the overall geoeffectiveness is smaller for limb halos. This study also revealed that one of the major geomagnetic storm was caused by a halo CME originating very close to the east limb, but the CME was extremely fast. Finally, most of the large geomagnetic storms are caused by disk halos, but a significant number (~9%) are caused by limb halos.

**Acknowledgments**

Data for Figures 1, 3 and 5 were obtained from NASA's OMNIweb (http://omniweb.gsfc.nasa.gov). Data for Figures 2 and 4 were obtained from the SOHO/LASCO catalog (http://cdaw.gsfc.nasa.gov) and from NOAA's GOES satellite flare listing. We acknowledge these data sources.